\documentclass[onecolumn,amssymb,aps]{revtex4}

\usepackage{algorithm}
\usepackage{algorithmic}

\newcommand{\rx}[1]{\ensuremath{\lceil #1 \rfloor}}
\newcommand{\rxd}[1]{\texttt{.\{0,} \ensuremath{#1} \texttt{\}}}

\makeatletter
\global\@specialpagefalse
\def\@oddhead{{\textit{Constans, P.}} \hfill \textit{Approximate retrieval } \thepage}
\let\@evenhead\@oddhead
\makeatother

\begin{document}

\title{Approximate textual retrieval}

\author{Pere Constans}
\email{constans@molspaces.com}
% \homepage{http://www.molspaces.com/}
\affiliation{Banyoles, May 2007}

\begin{abstract}
An approximate textual retrieval algorithm for searching sources with high
levels of defects is presented. It considers splitting the words in a query into
two overlapping segments and subsequently building composite regular expressions
from interlacing subsets of the segments. This procedure reduces the
probability of missed occurrences due to source defects, yet diminishes the
retrieval of irrelevant, non-contextual occurrences.
\end{abstract}
\maketitle

\section{Introduction}

Errors in electronic texts too often hinder the complete retrieval of the
intended search. Several approximate string matching or fault-tolerant
techniques have been devised to minorate their impact
\cite{Zobel95p331,Navarro01p31,Wang05p717,Esser05}. Albeit the variate number of
existing methods, which particularize specific retrieval needs, are all based on
word or character insertions, deletions and substitutions, which are performed
within a prescribed threshold of string similarity.

This article presents an algorithm for approximate string matching, suitable for
searching sources with high levels of defects. Concretely, it is devised to
search collections of scientific texts which are often encoded in electronic
formats that were originally created for printing and screen presentations.
Furthermore, earlier texts are recompiled through error-prone, optical character
recognition techniques.

Query strings are first split into words. Words are then divided into two
possibly overlapping segments, a \textit{prefix} and a \textit{suffix}.
Interlaced subsets are finally picked up from the ordered set of segments to
form a composite regular expression. On one hand, the composite regular
expression notably reduces the probability of missing a hit due to uniformly
distributed errors in the document. On the other hand, word segmentation and
query sectioning might lead to unveil hidden words perhaps irrelevant
to the query context.

Eliding parts of words or sentences, such as 'telephone' being set to 'phone' or
'zoological gardens' to 'zoo', and morphology derivations preserve in many
instances the semantics of the context \cite{Stockwellp}. The proposed algorithm
partitions words into their morphological constituents. The \textit{prefix}
segments embrace prefix and root; the \textit{suffix} segments, the root and
suffix. This partition gives longer segments and therefore reduces the
probability of irrelevant retrievals. The interlaced query sectioning and
inter-segment gap lengths are interrelated parameters in the algorithm. They
refer, intuitively, to an attention and resolution window within which the
documents are scanned. 

In the end, the extra computational effort that is necessary to reduce the
probability of missing a hit pays off when additional, related hits are
retrieved as well.

\section{Approximate textual retrieval algorithm}

Let $T$ be a text document constituted by a sequence $t_1 t_2 ...$ of words,
which, in turn, are sequences of characters over an alphabet $\Sigma$. Let the
query $\mathcal{Q}$ on $T$ for the word pattern $Q = q_1 q_2 ... q_n$ be defined
as the Boolean function
\begin{equation}
\mathcal{Q}_{T[j]} = [q_1 \sim t_{j+1}] \wedge [q_2 \sim t_{j+2}] \wedge ... \wedge [q_i \sim t_{j+i}] \wedge ... \wedge [q_n \sim t_{j+n}].
\end{equation}
The retrieval of pattern $Q$ from $T$ is then the set of positions $j$ for which
the query $\mathcal{Q}$ is true. 

If the word similarity relationships $[q_i \sim t_{j+i}]$ are set to equalities,
$[q_i = t_{j+i}]$, the probability of missing one pattern occurrence due to
uniformly distributed errors in $T$ is proportional to the length of $Q$. On the
other hand, the number of occurrences of $Q$ is proportional to the length of
$T$, whenever $T$ is a random text.

\subsection{Approximate composite queries}

Let the words in $Q$ be split into two segments, $q^p$ and $q^s$, such that
\begin{equation}
q = q^p \cup q^s,
\end{equation}
and let the word similarities $[q \sim t]$ be set to $[q^p \subset t]$
and $[q^s \subset t]$, meaning that $q$ is similar to $t$ if $q^p$ or $q^s$ are
segments of $t$. Furthermore, let $Q$ be sectioned into $b$ interlaced blocks to
build component queries. For $b = 2$, the two component queries are
\begin{eqnarray*}
\mathcal{Q}^p &=& [q^p_1 \subset t_{j+1}] \wedge [q^p_2 \subset t_{j+2}] \wedge ... \wedge [q^p_n \subset t_{j+n}] \\
\mathcal{Q}^s &=& [q^s_1 \subset t_{j+1}] \wedge [q^s_2 \subset t_{j+2}] \wedge ... \wedge [q^s_n \subset t_{j+n}]
\end{eqnarray*}
and the composite query $\tilde{\mathcal{Q}}$ is
\begin{equation}
\tilde{\mathcal{Q}} = \mathcal{Q}^p \cup \mathcal{Q}^s.
\end{equation}

More generally, being $R$ the relabeled sequence $Q$ of words
\begin{equation}
R = r_1 r_2 ... r_m = q^p_1 q^s_1 q^p_2 q^s_2 ... q^p_n q^s_n,
\end{equation}
the component queries are
\begin{eqnarray*}
\mathcal{R}^1 &=& \rx{r_1} \wedge \rx{\Sigma_{d1,1+b}} \wedge \rx{r_{1+b}} \wedge \rx{\Sigma_{d1+b,1+2b}} \wedge ... \wedge \rx{r_{1+\lceil m / b\rceil b - b}} \\
\mathcal{R}^2 &=& \rx{r_2} \wedge \rx{\Sigma_{d2,2+b}} \wedge \rx{r_{2+b}} \wedge \rx{\Sigma_{d2+b,2+2b}} \wedge ... \wedge \rx{r_{2+\lceil m / b\rceil b - b}}\\
&\vdots& \\
\mathcal{R}^b &=& \rx{r_b} \wedge \rx{\Sigma_{db,2b}} \wedge \rx{r_{2b}} \wedge \rx{\Sigma_{d2b,3b}} \wedge ... \wedge \rx{r_{\lfloor m / b\rfloor b}}.
\end{eqnarray*}

The approximate composite query $\tilde{\mathcal{R}}$ derived from
$\mathcal{Q}$ is then the union
\begin{equation} \tilde{\mathcal{R}} = \bigcup^{b}_{k = 1} \mathcal{R}^k. 
\end{equation}
In fact, $\tilde{\mathcal{R}}$ is an alternated regular expression. Notation
$\rx{\cdot}$ indicates match on $T$. $\rx{\Sigma_n}$ denotes match any segment
of characters in alphabet $\Sigma$ whose length $l$ is $0 \leq l \leq n$, and
$d_{i,i'}$ is the distance in characters from the last position of word $i$ to
the begin of word $i'$.

By construction, any component query $\mathcal{R}^k$ will match $Q$ in $T$
provided $\mathcal{Q}$ does. Their probabilities of missing one occurrence due
to random errors, $p_k$, are approximately equal to the one that $\mathcal{Q}$
has, divided by $b$. For the approximate composite query $\tilde{\mathcal{R}}$,
however, such probability is notably reduced, being of the order of $p^b$.

The expected number of matches that a regular expression of the form of
$\mathcal{R}^k$ will find in a random text has been analyzed by Flajolet,
Szpankowski and Vall{\'e}e \cite{Flajolet06p147}. If $\Omega_Q$ counts the
occurrences of pattern $Q$ in $T$, the expectation $E[\Omega_Q]$ is
approximately
\begin{equation}
E[\Omega_Q] = l_T {\textstyle \prod_i d_{ii'}} P(Q),
\label{retrieval}
\end{equation}
with $l_T$ being the length of $T$, $d_{ii'}$ the subpattern distances, and $P(Q)$
the probability of $Q$. The expectation for a composite expression
$\tilde{\mathcal{R}}$ is, therefore, approximately $b$ times $E[\Omega_Q]$.

\subsection{The algorithm}

As it has been implemented, the algorithm distinguishes two particular cases,
one for single and the other for multiple word queries. Since the number of
blocks $b$ cannot be greater than one plus the number of words, and since is
$b$ what permits escaping source defects, a word having errors in the segment
$q^p \cap q^s$ could not be matched. Furthermore, for a word without a clear
morphological partitioning, $q^p \cap q^s$ is equal to $q$. This case,
therefore, is treated separately, by considering that one single word can have
as much one single error, placed anywhere, but extending to no more than two
contiguous characters. This is a simple application of the insertion, deletion,
substitution paradigm. For the sake of completeness this case is also
included here.

The pseudo-codes for the two cases are listed in Algorithm \ref{algorithm_mw} and
\ref{algorithm_sw}, for the multiple and single word cases, respectively. They
are implemented in the \textsc{cb2Bib} program in version 0.8.2
\cite{Constans07}.

\begin{algorithm}
\caption{Approximate composite queries}
\label{algorithm_mw}
\begin{algorithmic}[1]

\FORALL{$q \in Q | l_q \ge 3$}
\STATE Split $q$ into $q^p$ and $q^s$ with $q = q^p \cup q^s$
\STATE $R \leftarrow R \cup q^p \cup q^s$
\ENDFOR

\FOR{$i = 1$ to $b$}
\STATE $\tilde{\mathcal{R}} \leftarrow \tilde{\mathcal{R}} \cup 
\rx{r_i\rxd{d_{i,i+b}}r_{i+b}\rxd{d_{i+b,i+ib}} \;\ldots\; r_{i+\lceil m / b\rceil b - b}}$
\ENDFOR

\end{algorithmic}
\end{algorithm}

\begin{algorithm}
\caption{Approximate single word matching}
\label{algorithm_sw}
\begin{algorithmic}[1]

\IF{$l_q < 3$}
\STATE $\tilde{\mathcal{R}} \leftarrow q$
\STATE return
\ENDIF

\FOR{$i = 1$ to $l_q$}
\STATE $\tilde{\mathcal{R}} \leftarrow \tilde{\mathcal{R}} \cup \rx{q[1:i-1]\rxd{2}q[i+1:l_q]}$
\ENDFOR

\end{algorithmic}
\end{algorithm}

\subsection{Remarks}

\begin{description}

\item[Word partition.]

The (approximate) partitioning of words into morphological parts is performed
using a look-up table composed of 1630 prefixes and 1133 suffixes. The listed
affixes also include combinations of them. In this manner, \textit{quant.ize.d},
for instance, will show its root \textit{quant} in the prefix+root portion, as
it will be shown by the related forms \textit{quant.ization} or
\textit{quant.um}. The word \textit{quantized} is therefore split into
\textit{quant} and \textit{quantized}. This produces longer forms that lower the
probability $P(Q)$ in equation \ref{retrieval}, and, hence, the chance of
unrelated occurrences.

\item[Interlacing blocks.]

The number of blocks $b$ expresses the portion of the query used by a composite
regular expression to scan the sources, being 
\begin{equation}
b = \min [b_{max}, 1 + 100 / \mathrm{percentScan}].
\end{equation}
The maximum number of blocks, $b_{max}$ is $\frac{1}{2}m$, or simply, the number
of words $n$.

\item[Misses and recall tradeoff.]

Besides setting the number of interlacing blocks, establishing appropriate gap
distances $d_{i,i'}$ is relevant regarding the tradeoff between missing
occurrences and overwhelming with unrelated ones. These two tuning parameters
are interrelated, being the minimum allowable distances dependent on the number
of blocks $b$. High values of $b$, or low percent scanning, greatly reduce the
probability of misses, but they increase the value of the product of distances
$d_{ii'}$ in equation \ref{retrieval}. In the current implementation, and for
the examples given in this work, the percent scanning has been set to 50\%.
Distances preceding high frequency words, i. e., words with four or less
characters, are set to three times their minimum allowable value. In the other
cases, they are set to either twenty times the difference $i' - i$, or three
times the allowable minimum, which ever is greater. This convention is
appropriate for searching a personal collection, where hits are hardly
irrelevant, due to its reduced and selected nature.

\end{description}

\subsection{Examples}

Two detailed examples are given to illustrate the algorithm for the cases of
single and multiple word queries. The queries are performed on the set of
articles cited in this work. Bold face font is used to highlight matched string
segments.

\subsubsection{Single word matching}

This example is taken from the work of Wang, Li, Cai, and Chen
\cite{Wang05p717}, on approximate string matching in biomedical text retrieval.
The search for 'chinensis' yields the regular expression:
\begin{small}
\begin{verbatim}
 (?:c(?:hinensi|hinen.{0,2}s|hine.{0,2}is|hin.{0,2}sis|hi.{0,2}nsis|h.{0,2}ensis|.{0,2}nensis)|hinensis)
\end{verbatim}
\end{small}
and produces two hits,

\begin{quotation}
\begin{small}
\textit{
$\bullet$ ...or -icus. Thus the name of 'Bupleurum \textbf{chinens}e' is
incorrect and the correct name is ``Bupleurum \textbf{chinensis}'' as shown in
Table 1. There are also... $\bullet$ ...II Grammatical error 86.6 Bupleurum
\textbf{chinens}e Bupleurum \textbf{chinensis} Collection II Grammatical error
89.5... $\bullet$ ...alba Collection II 29 36 Bupleurum \textbf{chinens}e
Collection II 23 28 Cinnamomum... $\bullet$ ...sachalinense Phellodendron
\textbf{chinens}e 84.2 Salvia przewalskii Sabina...
}
\end{small}
\end{quotation}
from reference \cite{Wang05p717}, matching the two spellings of the herb, and also,
\begin{quotation}
\begin{small}
\textit{
$\bullet$ ...tenths of seconds per megabyte. Our ma\textbf{chine is} a Sun UltraSparc-1 with 167 MHz and... 
}
\end{small}
\end{quotation}
from reference \cite{Navarro01p31}, and clearly not relevant.

\subsubsection{Composite queries}

The search for 'Aproximate textual retrieval' --note typo-- gives the word
segments \textit{Aproxim}, \textit{roximate}, \textit{textu}, \textit{textual},
\textit{retriev}, and \textit{rieval}, and the three-component regular
expression:
\begin{small}
\begin{verbatim}
 Aproxim.{0,60}textual
 roximate.{0,60}retriev
 textu.{0,60}rieval
\end{verbatim}
\end{small}
alternated as
\begin{small}
\begin{verbatim}
 (?:Aproxim.{0,60}textual|roximate.{0,60}retriev|textu.{0,60}rieval)
\end{verbatim}
\end{small}

It retrives the following texts,

\begin{quotation}
\begin{small}
\textit{
$\bullet$ ...the results above show that, for app\textbf{roximate matching,
they have speed and retriev}al effectiveness similar to that of 3...
}
\end{small}
\end{quotation}

\begin{quotation}
\begin{small}
\textit{
$\bullet$ ...references about the relation of app\textbf{roximate string
matching and information retriev}al are Wagner and Fisher [1974... $\bullet$
...2000. Blockaddressing indices for ap\textbf{proximate text retriev}al. J. Am.
Soc. Inf. Sci. (JASIS) 51... $\bullet$ ...SCHULMAN, E. 1997. Applications of
app\textbf{roximate word matching in information retriev}al. In Proceedings of
the 6th ACM...
}
\end{small}
\end{quotation}

\begin{quotation}
\begin{small}
\textit{
$\bullet$ ...Assessment of app\textbf{roximate string matching in a biomedical
text retriev}al problem J.F. Wang,a, Z.R. Lia,b , C... 
}
\end{small}
\end{quotation}

\begin{quotation}
\begin{small}
\textit{
$\bullet$ ...Keywords Fuzzy matching, app\textbf{roximate information
retrie}val, fault-tolerant fulltext search, q... $\bullet$ ...metric, used by
most available app\textbf{roximate text retriev}al algorithms, is not
appropriate when...
}
\end{small}
\end{quotation}
from the references \cite{Zobel95p331}, \cite{Navarro01p31}, \cite{Wang05p717},
and \cite{Esser05}, respectively.

Note that the three words in the query appear in two, and only two, component
expressions. Therefore, if the segment \textit{Approximate textual retrieval}
had been in the texts, the occurrence would have certainly been retrieved,
provided that the errors did not extend to more than one of the three words.

\section{Acknowledgment}

I am grateful to S. Vega for the careful, non-approximate reading of the manuscript.

\bibliographystyle{unsrt}
\bibliography{algorithms,pdfsearch}

\end{document}